\documentclass[iop]{emulateapj}
\usepackage{minipage}

\newcommand\kms{\ifmmode{\rm km\thinspace s^{-1}}\else km\thinspace s$^{-1}$\fi}
\newcommand\vstar{V506~Oph}
\newcommand\actaa{Acta Astron.}

\begin{document}
%\linenumbers

\submitted{Accepted for publication in The Astrophysical Journal}

\title{Absolute dimensions of the early F-type eclipsing binary V506~Ophiuchi}

\author{
Guillermo Torres\altaffilmark{1},
Claud H.\ Sandberg Lacy\altaffilmark{2},
Francis C. Fekel\altaffilmark{3}, and
Matthew W.\ Muterspaugh\altaffilmark{3,4}
}

\altaffiltext{1}{Center for Astrophysics \textbar\ Harvard \&
  Smithsonian, 60 Garden Street, Cambridge, MA 02138, USA;
  gtorres@cfa.harvard.edu}

\altaffiltext{2}{Physics Department, University of Arkansas,
  Fayetteville, AR 72701, USA}

\altaffiltext{3}{Center of Excellence in Information Systems,
  Tennessee State University, Nashville, TN 37209, USA}

\altaffiltext{4}{College of Life and Physical Sciences, Tennessee
  State University, Nashville, TN 37209, USA}

\begin{abstract}
We report extensive differential $V$-band photometry and
high-resolution spectroscopic observations of the early F-type,
1.06-day detached eclipsing binary \vstar. The observations along with
times of minimum light from the literature are used to derive a very
precise ephemeris and the physical properties for the components, with
the absolute masses and radii being determined to 0.7\% or better.
The masses are $1.4153 \pm 0.0100~M_{\sun}$ and
$1.4023 \pm 0.0094~M_{\sun}$ for the primary and secondary, the radii
are $1.725 \pm 0.010~R_{\sun}$ and $1.692 \pm 0.012~R_{\sun}$, and
the effective temperatures $6840 \pm 150$~K and $6780 \pm 110$~K,
respectively. The orbit is circular and the stars are rotating
synchronously. The accuracy of the radii and temperatures is supported
by the resulting distance estimate of $564 \pm 30$~pc, in excellent
agreement with the value implied by the trigonometric parallax listed
in the {\it Gaia\/}/DR2 catalog.  Current stellar evolution models
from the MIST series for a composition of ${\rm [Fe/H]} = -0.04$ match
the properties of both stars in \vstar\ very well at an age of
1.83~Gyr, and indicate they are halfway through their core
hydrogen-burning phase.
\end{abstract}


%%%%%%%%%%%%%%%%%%%%%%%%%%%%%%%%%%%%%%%%%%%%%%%%%%%%%%%%%%%%%%%%%%%%%%
\section{Introduction}
\label{sec:introduction}
%%%%%%%%%%%%%%%%%%%%%%%%%%%%%%%%%%%%%%%%%%%%%%%%%%%%%%%%%%%%%%%%%%%%%%

The variability of \vstar\ (TYC~993-1631-1, {\it Gaia}/DR2
4486661994344201344, $V = 11.1$, SpT \ion{F1}{5}) was discovered
photographically by \cite{Hoffmeister:1935}, who classified it as an
Algol-type eclipsing system. The binary orbital period of 1.06 days
was first established by \cite{Soloviev:1937}. Aside from the many
times of minimum light measured since, CCD light curves have been
reported ocassionally in the more recent literature
\citep{Pojmanski:2004, Lapham:2007, Kochanek:2017}, sometimes only in
graphical form, but there is no detailed study of the system as yet.

Here we report extensive new photometric observations of \vstar\ as
well as radial-velocity measurements, which we combine to determine
the physical properties of the system for the first time. The
spectroscopic observations and velocity measurements are presented in
Section~\ref{sec:spectroscopy}.  In Section~\ref{sec:specorbit} we
combine them with times of minimum light from the literature to derive
an accurate linear ephemeris as well as the spectroscopic elements.
The photometric observations are reported in
Section~\ref{sec:photometry}, and subjected to a detailed light curve
analysis in Section~\ref{sec:lightcurves}. The physical properties of
the stars, derived in Section~\ref{sec:dimensions}, are then compared
with predictions from recent stellar evolution models in
Section~\ref{sec:models}. Final remarks are given in
Section~\ref{sec:discussion}.

%%%%%%%%%%%%%%%%%%%%%%%%%%%%%%%%%%%%%%%%%%%%%%%%%%%%%%%%%%%%%%%%%%%%%%
\section{Spectroscopy}
\label{sec:spectroscopy}
%%%%%%%%%%%%%%%%%%%%%%%%%%%%%%%%%%%%%%%%%%%%%%%%%%%%%%%%%%%%%%%%%%%%%%

\vstar\ was observed spectroscopically with two different instruments.
Between 2010 May and 2017 February we monitored the binary with the
Center for Astrophysics \textbar\ Harvard \& Smithsonian (CfA)
Tillinghast Reflector Echelle Spectrograph
\citep[TRES;][]{Szentgyorgyi:2007, Furesz:2008} attached to the 1.5m
Tillinghast reflector at the Fred L.\ Whipple Observatory on Mount
Hopkins, Arizona. This bench-mounted, fiber-fed instrument delivers
spectra with a resolving power of $R \approx 44,000$ covering the
wavelength range 3900--9100~\AA\ in 51 orders. We gathered 48 spectra
with signal-to-noise ratios (S/N) near the \ion{Mg}{1}~b triplet
(5187~\AA) ranging from 21 to 74 per resolution element of
6.8~\kms. Wavelength calibrations relied on exposures of a
Thorium-Argon lamp taken before and after each science frame, and the
reductions were performed with a dedicated pipeline.

Radial velocities from the CfA spectra were measured with the
two-dimensional cross-correlation technique {\tt TODCOR}
\citep{Zucker:1994}. Templates appropriate for each star were taken
from a library of pre-computed synthetic spectra based on model
atmospheres by R.\ L.\ Kurucz \citep[see][]{Nordstrom:1994,
  Latham:2002}. For this analysis we used only the 100~\AA\ wide order
centered on the \ion{Mg}{1}~b triplet, as previous experience
indicates it contains most of the velocity information and because our
synthetic templates are limited in coverage to a narrow region
surrounding that feature. We selected the best templates by running
grids of cross-correlations over wide ranges in effective temperature
($T_{\rm eff}$) and rotational broadening ($v \sin i$ when seen in
projection), at fixed solar metallicity and values of the surface
gravity $\log g = 4.0$, close to our final determinations in
Section~\ref{sec:dimensions}. Following \cite{Torres:2002} we selected
the template parameters giving the highest cross-correlation value
averaged over all observations, with weights set by the strength of
each exposure. In this way we estimated the temperatures to be $6840
\pm 150$~K and $6860 \pm 150$~K for the primary (the marginally more
massive star) and secondary, which are the same within their
uncertainties. They correspond approximately to spectral type F1.  The
uncertainties are based on the scatter from the individual spectra,
with an extra 100~K added in quadrature, to be conservative. The
rotational velocities were determined to be $80 \pm 3$~\kms\ for both
stars. Thus the spectroscopic properties are essentially identical.
The light ratio at the mean wavelength of our observations
\citep[see][]{Zucker:1994} was found to be $\ell_2/\ell_1 = 0.96 \pm
0.03$. The resulting radial velocities in the heliocentric frame are
listed in Table~\ref{tab:rvs_cfa} along with their uncertainties.

\setlength{\tabcolsep}{3pt}  % tighten to make table fit in one column
\begin{deluxetable}{crcrcc}
\tablewidth{0pc}
\tablecaption{Heliocentric radial velocity measurements of
  \vstar\ from CfA. \label{tab:rvs_cfa}}
\tablehead{
\colhead{HJD} &
\colhead{$RV_1$} &
\colhead{$\sigma_1$} &
\colhead{$RV_2$} &
\colhead{$\sigma_2$} &
\colhead{Orbital}
\\
\colhead{(2,400,000+)} &
\colhead{(\kms)} &
\colhead{(\kms)} &
\colhead{(\kms)} &
\colhead{(\kms)} &
\colhead{Phase}
}
\startdata
     55338.8444 &     122.23 &     3.66 & $-$134.06 &     3.90 &    0.0881 \\ 
     55347.7766 &  $-$149.05 &     2.70 &    146.47 &     2.88 &    0.5113 \\ 
     55369.9595 &  $-$124.82 &     4.73 &    127.38 &     5.05 &    0.4302 \\ 
     55381.6768 &  $-$148.08 &     3.01 &    142.72 &     3.21 &    0.4798 \\ 
     55640.9437 &     138.89 &     2.73 & $-$142.29 &     2.91 &    0.9725 \\ 
     55758.7037 &     140.27 &     2.25 & $-$148.18 &     2.41 &    0.0220 \\ 
     56019.0159 &  $-$147.43 &     4.21 &    148.12 &     4.49 &    0.5005 \\ 
     56027.9888 &     139.07 &     2.50 & $-$147.22 &     2.67 &    0.9621 \\ 
     56058.8175 &     141.14 &     3.92 & $-$146.90 &     4.18 &    0.0341 \\ 
     56074.7773 &     119.63 &     1.80 & $-$128.91 &     1.92 &    0.0844 \\ 
     56135.7569 &  $-$124.63 &     2.70 &    128.16 &     2.88 &    0.5891 \\ 
     56205.5940 &  $-$143.66 &     2.95 &    135.05 &     3.15 &    0.4466 \\ 
     56351.0042 &  $-$137.24 &     2.81 &    133.00 &     3.00 &    0.5707 \\ 
     56375.9216 &     129.97 &     2.84 & $-$133.45 &     3.03 &    0.0682 \\ 
     56401.9332 &  $-$121.29 &     2.82 &    116.96 &     3.02 &    0.5975 \\ 
     56404.9811 &  $-$150.90 &     2.42 &    139.02 &     2.58 &    0.4718 \\ 
     56435.8296 &  $-$141.08 &     2.18 &    133.32 &     2.33 &    0.5624 \\ 
     56436.8014 &  $-$150.41 &     2.44 &    146.09 &     2.60 &    0.4788 \\ 
     56460.7272 &     139.48 &     4.62 & $-$147.98 &     4.93 &    0.0412 \\ 
     56502.6813 &  $-$120.78 &     3.66 &    119.43 &     3.90 &    0.6046 \\ 
     56554.6252 &  $-$125.01 &     1.98 &    122.39 &     2.11 &    0.5885 \\ 
     56739.0051 &  $-$154.96 &     3.74 &    142.25 &     3.99 &    0.4616 \\ 
     56761.9152 &     135.28 &     4.07 & $-$137.51 &     4.34 &    0.0662 \\ 
     56795.7617 &     142.51 &     3.36 & $-$148.01 &     3.59 &    0.9840 \\ 
     56816.9153 &     132.73 &     3.14 & $-$138.07 &     3.36 &    0.9322 \\ 
     56824.8427 &  $-$130.78 &     2.82 &    124.13 &     3.01 &    0.4078 \\ 
     57063.0212 &     143.69 &     3.75 & $-$146.77 &     4.00 &    0.0139 \\ 
     57086.9575 &  $-$127.83 &     2.65 &    125.97 &     2.83 &    0.5862 \\ 
     57089.0347 &  $-$147.38 &     4.46 &    140.41 &     4.76 &    0.5450 \\ 
     57091.0266 &  $-$131.47 &     3.32 &    123.07 &     3.55 &    0.4234 \\ 
     57114.0118 &     118.77 &     4.23 & $-$122.62 &     4.52 &    0.0988 \\ 
     57118.0005 &      85.33 &     3.24 & $-$102.77 &     3.46 &    0.8602 \\ 
     57141.9453 &  $-$144.11 &     3.95 &    135.13 &     4.22 &    0.4405 \\ 
     57168.9740 &     127.31 &     3.60 & $-$138.77 &     3.84 &    0.9290 \\ 
     57207.7905 &  $-$145.52 &     3.50 &    140.23 &     3.74 &    0.5336 \\ 
     57291.6177 &  $-$123.53 &     3.82 &    114.85 &     4.08 &    0.5840 \\ 
     57443.0342 &  $-$108.50 &     2.83 &     94.27 &     3.02 &    0.3721 \\ 
     57447.0052 &     105.52 &     2.49 & $-$113.92 &     2.66 &    0.1168 \\ 
     57472.9373 &  $-$130.44 &     3.86 &    132.20 &     4.12 &    0.5712 \\ 
     57498.8922 &     136.32 &     2.60 & $-$141.82 &     2.77 &    0.0471 \\ 
     57514.8946 &      93.96 &     4.26 &  $-$99.61 &     4.55 &    0.1376 \\ 
     57523.9009 &  $-$106.32 &     3.84 &     99.23 &     4.09 &    0.6307 \\ 
     57534.8209 &     125.13 &     3.13 & $-$134.26 &     3.34 &    0.9284 \\ 
     57558.6875 &  $-$141.76 &     3.13 &    133.24 &     3.34 &    0.4350 \\ 
     57581.6919 &      99.09 &     2.36 & $-$108.41 &     2.52 &    0.1285 \\ 
     57585.7499 &     134.28 &     3.77 & $-$143.35 &     4.03 &    0.9552 \\ 
     57598.6716 &      87.23 &     2.89 & $-$100.39 &     3.09 &    0.1406 \\ 
     57807.0186 &  $-$112.31 &     3.77 &    101.98 &     4.03 &    0.6151 %\\ [-1.5ex]
\enddata
\tablecomments{Phases are calculated from the reference time of primary eclipse in Table~\ref{tab:specorbit}.}
\end{deluxetable}
\setlength{\tabcolsep}{6pt}  % back to default separation

\vstar\ was also observed at Fairborn Observatory in southeast Arizona
near Washington Camp, between 2012 February and 2018 May. For this we
used the Tennessee State University 2m Astronomical Spectroscopic
Telescope (AST) and a fiber fed echelle spectrograph
\citep{Eaton:2007}. The detector was a Fairchild 486 CCD having
4K$\times$4K pixels 15~$\mu$m in size, which results in echelle
spectra that have 48 orders and cover a wavelength range of
3800--8260~\AA\ \citep{Fekel:2013}. Because of the faintness of the
system, we used a fiber diameter that produced a spectral resolution
of 0.4~\AA, but even so, given the weakness and very significant line
broadening of the features, many of the spectra did not have a high
enough S/N to provide meaningful results. However, we were able to
obtain useful velocity measurements from 17 AST spectra that had a
resolving power of 15000 at 6000~\AA\ and an average S/N of about 40.

A description of the general radial velocity reduction of the Fairborn
AST spectra has been given by \cite{Fekel:2009}. In particular for
\vstar\ we used a solar line list that consisted of 168 mostly neutral
Fe lines that cover a wavelength range of 4920--7100~\AA.  The
individual lines were fitted with a rotational broadening function
\citep{Lacy:2011}.  Unpublished velocities of several IAU solar-type
radial velocity standards show that velocities obtained with our
Fairchild CCD have a $-0.6~\kms$ offset relative to the velocities of
\cite{Scarfe:2010}. Thus, 0.6~\kms\ has been added to each velocity.
We list these measurements in Table~\ref{tab:rvs_fek}. We estimate the
uncertainties to be 3.2 and 2.6~\kms\ for the primary and secondary,
respectively, from the scatter of a preliminary spectroscopic orbital
solution.

\begin{deluxetable}{crrc}
\tablewidth{0pc}
\tablecaption{Heliocentric radial velocity measurements of
  \vstar\ from Fairborn Observatory. \label{tab:rvs_fek}}
\tablehead{
\colhead{HJD} &
\colhead{$RV_1$} &
\colhead{$RV_2$} &
\colhead{Orbital}
\\
\colhead{(2,400,000+)} &
\colhead{(\kms)} &
\colhead{(\kms)} &
\colhead{Phase}
}
\startdata
     55976.9665 &     82.80 &  $-$92.80 &    0.5980 \\ 
     56419.9033 & $-$144.60 &    137.40 &    0.2945 \\ 
     56454.8887 & $-$147.90 &    140.10 &    0.2863 \\ 
     56565.6922 &    138.30 & $-$149.00 &    0.7757 \\ 
     56731.9907 &     80.20 &  $-$92.60 &    0.5979 \\ 
     56769.8765 & $-$140.30 &    130.50 &    0.3248 \\ 
     56799.8553 &     79.40 &  $-$94.00 &    0.5953 \\ 
     56902.7375 &     92.80 & $-$102.00 &    0.6149 \\ 
     57088.9577 & $-$150.00 &    142.40 &    0.2235 \\ 
     57174.7409 & $-$108.70 &     98.60 &    0.1184 \\ 
     57509.8563 & $-$123.30 &    113.10 &    0.1376 \\ 
     57541.8116 & $-$145.70 &    145.20 &    0.2720 \\ 
     57595.7453 & $-$110.80 &    104.20 &    0.1323 \\ 
     57653.6778 &    146.00 & $-$149.70 &    0.7636 \\ 
     57851.8666 &    120.80 & $-$126.50 &    0.6588 \\ 
     58003.6971 &    116.70 & $-$129.90 &    0.8374 \\ 
     58245.8184 & $-$129.50 &    121.30 &    0.1616 %\\ [-1.5ex]
\enddata

\tablecomments{Velocity uncertainties are esimated to be 3.2 and 2.6~\kms\ for the primary and secondary, respectively. Phases are calculated from the reference time of primary eclipse in Table~\ref{tab:specorbit}.}

\end{deluxetable}

Rotational broadening fits of the stellar lines in our 17 spectra
result in $v \sin i$ values of $81 \pm 3~\kms$ for both components.
From the same spectra the average equivalent width ratio of the
secondary to the primary, which should be equivalent to the light
ratio since the spectra appear to be very similar, is $\ell_2/\ell_1 =
0.96 \pm 0.03$, the same as obtained from the CfA spectra.

%%%%%%%%%%%%%%%%%%%%%%%%%%%%%%%%%%%%%%%%%%%%%%%%%%%%%%%%%%%%%%%%%%%%%%
\section{Times of minimum and spectroscopic orbit}
\label{sec:specorbit}
%%%%%%%%%%%%%%%%%%%%%%%%%%%%%%%%%%%%%%%%%%%%%%%%%%%%%%%%%%%%%%%%%%%%%%

Times of minimum light for \vstar\ have been recorded since 1928 by
photographic, visual, and photoelectric/CCD techniques. We collect all
176 measurements of which we are aware (84 for the primary and 92 for
the secondary) in Table~\ref{tab:minima}, with their uncertainties
when published.

\setlength{\tabcolsep}{2.5pt}  % tighten to make table fit in one column
\begin{deluxetable}{cccccrc}
\tablewidth{0pc}
\tablecaption{Times of minimum light for \vstar. \label{tab:minima}}
\tablehead{
\colhead{HJD} &
\colhead{$\sigma$} &
\colhead{} &
\colhead{} &
\colhead{} &
\colhead{$(O-C)$} &
\colhead{}
\\
\colhead{(2,400,000+)} &
\colhead{(days)} &
\colhead{Eclipse} &
\colhead{Type} &
\colhead{Year} &
\colhead{(days)} &
\colhead{Source}
}
\startdata
  25502.313 & \nodata & 2  &  pg &  1928.6990  &   0.01250  &  1 \\
  26068.578 & \nodata & 2  &  pg &  1930.2494  &   0.00928  &  1 \\
  26145.469 & \nodata & 1  &  pg &  1930.4599  &   0.01929  &  1 \\
  26592.424 & \nodata & 2  &  pg &  1931.6836  &   0.00415  &  1 \\
  26856.481 & \nodata & 2  &  pg &  1932.4065  &   0.01473  &  1 %\\ [-1.5ex]
\enddata

\tablecomments{The uncertainties in the second column are taken
  directly from the original publications. Scale factors for these
  errors determined from our joint solution with the spectroscopy are
  given in the text. The `Eclipse' column refers to the primary (1) or
  secondary (2) minimum. `Type' is ``pg'', ``v'', or ``pe'' for
  photographic, visual, or photoelectric/CCD observations. Sources
  are: (1)
  \url{https://www.bav-astro.eu/index.php/veroeffentlichungen/service-for-scientists/lkdb-engl};
  (2) \url{http://var2.astro.cz/ocgate/?lang=en}; (3)
  \cite{Lapham:2007}, with the unrealistically small formal
  uncertainties multiplied by 30; (4) \cite{Lacy:2007}. (This table is
  available in its entirety in machine-readable form.)}
\end{deluxetable}
\setlength{\tabcolsep}{6pt}  % back to default separation

\begin{figure}
\epsscale{1.15} \plotone{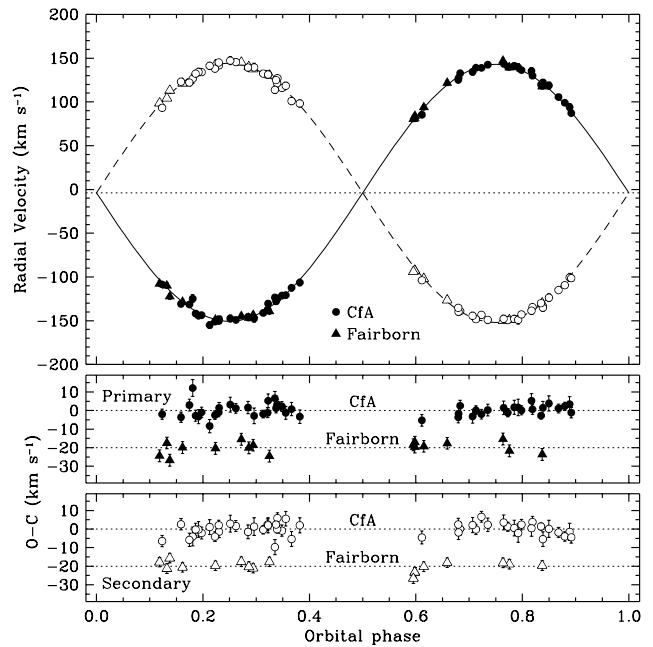}

\figcaption{Radial velocity observations along with our joint
  spectroscopic orbital solution incorporating the times of minimum
  light. The dotted line marks the center-of-mass velocity of the
  system.  Residuals are shown at the bottom, with the ones from
  Fairborn Observatory displaced vertically for clarity. Phases are
  counted from the reference time of primary
  eclipse in Table~\ref{tab:specorbit}. \label{fig:specorbit}}
\end{figure}

Independent spectroscopic orbital solutions from the CfA and Fairborn
velocities gave elements consistent with each other, except for a
minor difference in the center-of-mass velocities that is of no
consequence and is likely due to instrumental shifts. We therefore
combined these data sets. Furthermore, as the times of minimum light
spanning nearly 87 years constrain the ephemeris far better than our
radial velocities can, we used the two kinds of observations together
in a joint orbital solution to derive the final ephemeris and
spectroscopic elements simultaneously. For the times of minimum
without published uncertainties we adopted errors of 0.0175, 0.0146,
and 0.0035 days for the photographic, visual, and photoelectric/CCD
measurements, respectively, determined by iterations so as to achieve
reduced $\chi^2$ values near unity for each type of observation. In a
similar manner we determined appropriate scaling factors to be applied
to the published visual errors of 1.09 and 1.28 for the primary and
secondary measurements, and scale factors for the photoelectric/CCD
errors of 1.17 and 1.65. Initial fits allowing for separate epochs of
primary and secondary minumum showed no evidence of eccentricity, so
the final fit assumed none. We also allowed for possible velocity
offsets between the primary and secondary stars separately for the CfA
and Fairborn data ($\Delta {\rm RV}_{\rm CfA}$, $\Delta {\rm RV}_{\rm
  Fairborn}$), which in the case of the CfA data may result from
template mismatch. We additionally solved for a systematic offset
between the CfA and Fairborn velocity zero points ($\Delta {\rm RV}$),
to account for possible instrumental shifts as indicated above. The
results are listed in Table~\ref{tab:specorbit}, and shown graphically
in Figure~\ref{fig:specorbit} together with the observations and
residuals.

\begin{deluxetable}{lc}
\tablewidth{0pc}
\tablecaption{Spectroscopic orbital elements of \vstar. \label{tab:specorbit}}
\tablehead{
\colhead{~~~~~~~~~~~~Parameter~~~~~~~~~~~~} &
\colhead{Value}
}
\startdata
\multicolumn{2}{c}{Adjusted elements} \\ [0.5ex]
\noalign{\hrule} \\ [-2ex]
$P$ (days)\dotfill                         & $1.060427381 \pm 0.000000024$ \\
$\gamma$ (\kms)\dotfill                    & $-3.88 \pm  0.43$\phs \\
$K_1$ (\kms)\dotfill                       & $146.76 \pm 0.44$\phn\phn \\
$K_2$ (\kms)\dotfill                       & $148.11 \pm 0.45$\phn\phn \\
Min~I (${\rm HJD}-2,\!400,\!000$)\dotfill      & $53,\!123.782733 \pm 0.000037$\phm{222,2} \\
$\Delta {\rm RV}_{\rm CfA}$ (\kms)\dotfill & $-0.99 \pm 0.65$\phs \\
$\Delta {\rm RV}_{\rm Fairborn}$ (\kms)\dotfill    & $-0.98 \pm 1.03$\phs \\
$\Delta {\rm RV}$ (\kms)\dotfill           & $+1.02 \pm 0.91$\phs \\ [0.5ex]
\noalign{\hrule} \\ [-1.5ex]
\multicolumn{2}{c}{Derived quantities} \\ [0.5ex]
\noalign{\hrule} \\ [-2ex]
$M_1 \sin^3 i$ ($M_{\sun}^{\rm N}$)\dotfill        & $1.4151 \pm 0.0096$ \\
$M_2 \sin^3 i$ ($M_{\sun}^{\rm N}$)\dotfill        & $1.4022 \pm 0.0094$ \\
$q \equiv M_2/M_1$\dotfill                 & $0.9909 \pm 0.0042$ \\
$a_1 \sin i$ ($10^6$ km)\dotfill         & $2.1401 \pm 0.0064$ \\
$a_2 \sin i$ ($10^6$ km)\dotfill                 & $2.1598 \pm 0.0066$ \\
$a \sin i$ ($R_{\sun}^{\rm N}$)\dotfill            & $6.181 \pm 0.013$ \\
CfA $\sigma_1$, $\sigma_2$ (\kms)\dotfill       & 2.91,~3.26 \\
CfA $N_{\rm RV,1}$, $N_{\rm RV,2}$\dotfill      & 48,~48 \\
Fairborn $\sigma_1$, $\sigma_2$ (\kms)\dotfill  & 3.20,~2.60 \\
Fairborn $N_{\rm RV,1}$, $N_{\rm RV,2}$\dotfill  & 17,~17 \\
$N_{\rm Min~I}$, $N_{\rm Min~II}$\dotfill       & 84,~92 %\\ [-1.5ex]
\enddata

\tablecomments{$\Delta_{\rm CfA}$ and $\Delta_{\rm Fairborn}$
  represent the primary minus secondary velocity offsets, and $\Delta
  {\rm RV}$ the global CfA minus Fairborn shift. The minimum masses
  and semimajor axis are expressed in units of the nominal solar mass
  and radius ($M_{\sun}^{\rm N}$, $R_{\sun}^{\rm N}$) as recommended
  by 2015 IAU Resolution B3 \citep[see][]{Prsa:2016}.}

\end{deluxetable}

%%%%%%%%%%%%%%%%%%%%%%%%%%%%%%%%%%%%%%%%%%%%%%%%%%%%%%%%%%%%%%%%%%%%%%
\section{Photometry}
\label{sec:photometry}
%%%%%%%%%%%%%%%%%%%%%%%%%%%%%%%%%%%%%%%%%%%%%%%%%%%%%%%%%%%%%%%%%%%%%%

Differential photometry of \vstar\ in the $V$ band was performed with
the URSA WebScope at the University of Arkansas at Fayetteville and
with the NFO WebScope near Silver City, New Mexico (see
\citealt{Lacy:2014} for technical details). \vstar\ (var) was measured
along with two nearby comparison stars (comp; TYC~993-762-1, $V =
11.30$, $B-V = 2.08$, and TYC~993-0780-1, $V = 10.78$, $B-V = 0.49$).
Differential magnitudes were measured with the application {\tt
  Measure} written by author Lacy.  The two comparison star fluxes
were combined and the differential magnitudes were calculated as
var$-$comps. We obtained 8345 URSA images between 2003 July and 2012
June on a total of 129 nights, and 7475 NFO images between 2005
January and 2013 June on a total of 234 nights.  Exposures were
120~sec long, and square photometric apertures with sizes of
30\arcsec\ and 22\arcsec\ were used for URSA and NFO,
respectively. The {\it Gaia}/DR2 catalog lists 7 nearby stars within
30\arcsec\ of \vstar, but they are all at least 8 magnitudes fainter
and therefore do not contaminate the photometry.

\begin{figure*}
\centering
\includegraphics[width=11cm,angle=270]{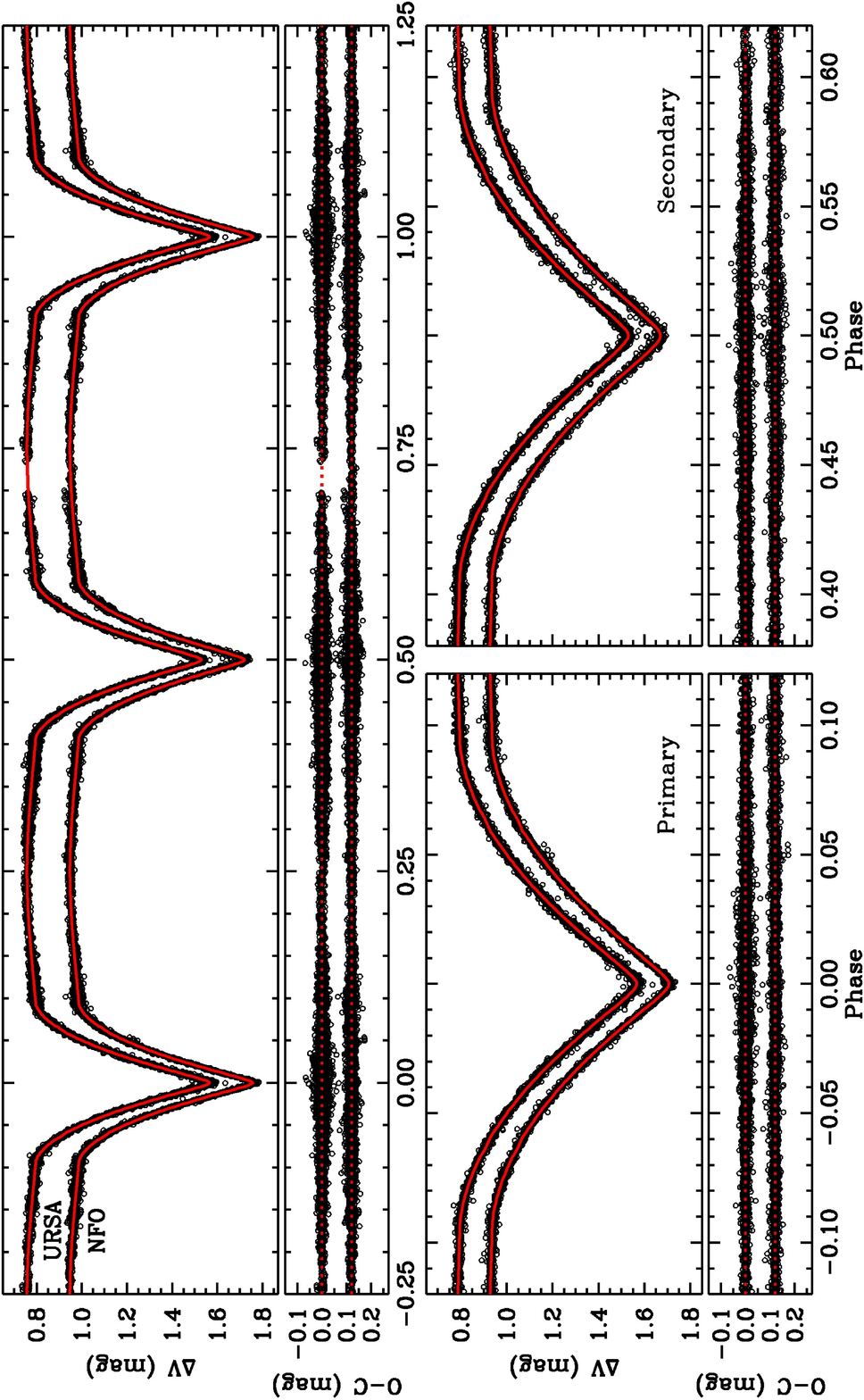}
\caption{Photometric observations of \vstar\ from the URSA and NFO
  telescopes, along with the residuals. NFO is displaced vertically
  for clarity. Enlargements of the minima are shown at the bottom. Our
  adopted model discussed in Section~\ref{sec:lightcurves} is
  overplotted. \label{fig:lightcurves}}
\end{figure*}

Examination of the raw data revealed that the NFO measurements suffer
from small systematic errors typically less than 0.02~mag, caused
by imprecise centering from night to night and variations in
responsivity across the field of view \citep[see][]{Lacy:2014}. We
corrected this by applying nightly offsets based on a preliminary
light curve solution using the URSA data alone, which shows no such
effects for \vstar. The full data sets are given in
Table~\ref{tab:ursa} (URSA) and Table~\ref{tab:nfo} (NFO, including
corrections).  The resultant light curves are displayed in
Figure~\ref{fig:lightcurves}.

\begin{table}
\centering
\hskip -10pt
\begin{minipage}[b]{0.38\linewidth}\centering
\caption{URSA observations of \\ \vstar.\label{tab:ursa}}
\begin{tabular}{cc}
\hline \\ [-2.5ex]
\hline \\ [-2.5ex]
HJD & $\Delta V$ \\
(2,400,000+) & (mag) \\ [+0.5ex]
\hline \\ [-2ex]
  52831.60573 &  1.211 \\
  52831.60763 &  1.232 \\
  52831.60954 &  1.250 \\
  52831.61143 &  1.305 \\
  52831.61329 &  1.296 \\ [0.5ex]
\hline \\ [-2ex]
\multicolumn{2}{p{\textwidth}}{{\bf Note}. --- (This table is available in its entirety in machine-readable form.)}
\end{tabular}
\end{minipage}
\hskip 25pt
\begin{minipage}[b]{0.38\linewidth}\centering
\caption{NFO observations of \\ \vstar.\label{tab:nfo}}
\begin{tabular}{cc}
\hline \\ [-2.5ex]
\hline \\ [-2.5ex]
HJD & $\Delta V$ \\
(2,400,000+) & (mag) \\ [+0.5ex]
\hline \\ [-2ex]
 53399.02069 &  0.960 \\
 53399.02460 &  0.923 \\
 53399.02849 &  0.911 \\
 53399.03236 &  0.884 \\
 53399.03625 &  0.862 \\ [0.5ex]
\hline \\ [-2ex]
\multicolumn{2}{p{\textwidth}}{{\bf Note}. --- (This table is available in its entirety in machine-readable form.)}
\end{tabular}
\end{minipage}
\end{table}

%%%%%%%%%%%%%%%%%%%%%%%%%%%%%%%%%%%%%%%%%%%%%%%%%%%%%%%%%%%%%%%%%%%%%%
\section{Light curve analysis}
\label{sec:lightcurves}
%%%%%%%%%%%%%%%%%%%%%%%%%%%%%%%%%%%%%%%%%%%%%%%%%%%%%%%%%%%%%%%%%%%%%%

The URSA and NFO photometry of \vstar\ was analyzed using version 2013
of the Wilson-Devinney {\tt LC\/} program \citep{Wilson:1971,
  Wilson:1979, Wilson:1990} called within a Markov Chain Monte Carlo
(MCMC) scheme. Our method of solution used the {\tt
  emcee\/}\footnote{\url http://dan.iel.fm/emcee~.} code of
\cite{Foreman-Mackey:2013}, which is a Python implementation of the
affine-invariant MCMC ensemble sampler proposed by
\cite{Goodman:2010}. We typically used 100 walkers and uniform priors
within suitable limits for all fitted quantities.

As the system is well detached we used the {\tt LC\/} program in mode
2, along with the option of simple reflection and synchronous rotation
of both components (see Section~\ref{sec:dimensions}).  The ephemeris
and mass ratio of the binary were held fixed at the values in
Table~\ref{tab:specorbit}, and the primary temperature was set to
6840~K (Section~\ref{sec:spectroscopy}). The main parameters of the
fit were the inclination angle $i$, the temperature of the secondary
$T_{\rm eff,2}$, the surface potentials $\Phi_1$ and $\Phi_2$, a phase
shift $\Delta \phi$, and the out-of-eclipse magnitude difference at
phase 0.25, $m_0$.  We assumed initial measurement errors for the URSA
and NFO observations of 0.02~mag, and a scale factor $f$ (with a
log-uniform prior) was included as an additional adjustable parameter,
which we solved for self-consistently and simultaneously with the
other parameters \citep[see][]{Gregory:2005}. Convergence of the
chains was checked visually, with the additional requirement of a
Gelman-Rubin statistic of 1.05 or smaller for each parameter.

The URSA and NFO data sets were initially analyzed separately. Tests
indicated the best results were obtained by solving also for the
linear limb-darkening coefficients of each star ($x_1$, $x_2$), as
well as the gravity-darkening exponents ($\beta_1$, $\beta_2$). More
complicated limb-darkening laws did not provide any improvement. The
albedos for both components were held fixed at a value of 0.5,
commonly adopted for convective stars, as experiments with other
values gave poorer results. No significant third light was detected,
consistent with the fact that the {\it Gaia}/DR2 catalog
\citep{Gaia:2018} lists no companions within the photometric apertures
that are bright enough to affect the light curves.

The independent URSA and NFO solutions gave similar results, so for
our final solution we solved both light curves together imposing a
common geometry as well as a single value of $T_{\rm eff,2}$ and the
limb- and gravity-darkening parameters for each star, for a total of
14 free parameters. The resulting light elements are presented in
Table~\ref{tab:lightcurves}, and the adopted model is shown in
Figure~\ref{fig:lightcurves} overlaid on the observations.

\begin{table}
\centering
\caption{Light curve elements of \vstar\ from our combined
  URSA+NFO solution. \label{tab:lightcurves}}
\begin{tabular*}{0.47\textwidth}{lcc}
\hline \\ [-2.5ex]
\hline \\ [-2.5ex]
~~~~~~~~~Parameter~~~~~~~~~ & Primary & Secondary \\ [+0.5ex]
\hline \\ [-2.5ex]
\multicolumn{3}{c}{Adjusted elements} \\ [0.5ex]
\noalign{\hrule} \\ [-2ex]
$i$ (deg)\dotfill             & \multicolumn{2}{c}{$89.27^{+0.18}_{-0.16}$} \\ [1ex]
$T_{\rm eff}$ (K)\dotfill   & 6840 (fixed) & $6781^{+110}_{-110}$ \\ [1ex]
$\Phi$\dotfill    & $4.629^{+0.019}_{-0.021}$ & $4.678^{+0.029}_{-0.027}$ \\ [1ex]
$x$\dotfill       & $0.462^{+0.048}_{-0.048}$ & $0.455^{+0.052}_{-0.052}$ \\ [1ex]
$\beta$\dotfill & $0.47^{+0.12}_{-0.13}$ & $0.42^{+0.16}_{-0.16}$ \\ [1ex]
$\Delta\phi_{\rm URSA}$\dotfill & \multicolumn{2}{c}{$-0.000029^{+0.000027}_{-0.000030}$} \\ [1ex]
$\Delta\phi_{\rm NFO}$\dotfill & \multicolumn{2}{c}{$-0.000039^{+0.000031}_{-0.000031}$} \\ [1ex]
$m_{\rm 0,URSA}$ (mag)\dotfill & \multicolumn{2}{c}{$0.75675^{+0.00057}_{-0.00054}$} \\ [1ex]
$m_{\rm 0,NFO}$ (mag)\dotfill & \multicolumn{2}{c}{$0.74756^{+0.00054}_{-0.00053}$} \\ [1ex]
$\ln f_{\rm URSA}$\dotfill  & \multicolumn{2}{c}{$-0.6172^{+0.0079}_{-0.0079}$} \\ [1ex]
$\ln f_{\rm NFO}$\dotfill  & \multicolumn{2}{c}{$-0.7513^{+0.0083}_{-0.0083}$} \\ [1ex]
\noalign{\hrule} \\ [-2.5ex]
\multicolumn{3}{c}{Derived quantities} \\ [0.5ex]
\noalign{\hrule} \\ [-2ex]
$r_{\rm pole}$\dotfill   &  $0.2723^{+0.0016}_{-0.0014}$ & $0.2673^{+0.0016}_{-0.0017}$ \\ [1ex]
$r_{\rm point}$\dotfill  &  $0.2911^{+0.0021}_{-0.0018}$ & $0.2848^{+0.0021}_{-0.0022}$ \\ [1ex]
$r_{\rm side}$\dotfill   &  $0.2780^{+0.0017}_{-0.0015}$ & $0.2726^{+0.0017}_{-0.0018}$ \\ [1ex]
$r_{\rm back}$\dotfill   &  $0.2863^{+0.0019}_{-0.0017}$ & $0.2804^{+0.0019}_{-0.0020}$ \\ [1ex]
$r_{\rm vol}$\dotfill    &  $0.2791^{+0.0017}_{-0.0015}$ & $0.2737^{+0.0017}_{-0.0018}$ \\ [1ex]
$r_1+r_2$\dotfill                &       \multicolumn{2}{c}{$0.5529^{+0.0010}_{-0.0010}$} \\ [1ex]
$r_2/r_1$\dotfill                &       \multicolumn{2}{c}{$0.981^{+0.011}_{-0.012}$} \\ [1ex]
$(\ell_2/\ell_1)_V$\dotfill      &       \multicolumn{2}{c}{$0.930^{+0.014}_{-0.015}$} \\ [1ex]
$\Delta T_{\rm eff}$ (K)\dotfill &       \multicolumn{2}{c}{$59^{+23}_{-24}$} \\ [1ex]
$\sigma_{\rm URSA}$, $\sigma_{\rm NFO}$ (mag)\dotfill & \multicolumn{2}{c}{0.01079,~0.00944} \\
$N_{\rm URSA}$, $N_{\rm NFO}$\dotfill   & \multicolumn{2}{c}{8345,~7475} \\ [0.5ex]
\hline \\ [-2ex]

\multicolumn{3}{p{.46\textwidth}}{{\bf Note}. --- The parameter values
listed correspond to the mode of the posterior distributions, and the
uncertainties are the 16\% and 84\% (1$\sigma$) credible intervals.}
\end{tabular*}
\end{table}

To guard against the possibility that the uncertainties returned by
our MCMC procedure are underestimated because of residual systematic
errors (i.e., time-correlated or ``red'' noise) in the NFO data, or
even in the URSA data, we carried out a residual permutation exercise
as described next. The light curve residuals from our adopted solution
were shifted by an arbitrary number of time steps (separately for URSA
and NFO) and added back into the model curve at each time of
observation (with wrap-around) to create synthetic data sets. We
subjected them to a new MCMC solution in which we simultaneously
perturbed the primary temperature, the mass ratio, and the albedos by
adding Gaussian noise with standard deviations equal to their reported
uncertainties for $T_{\rm eff,1}$ and $q$, and $\sigma = 0.2$ for the
albedos. We repeated this 50 times, and computed the scatter (standard
deviation) of the resulting distribution for each fitted and derived
parameter as a measure of the uncertainty caused by red noise.  We
then added this uncertainty and the internal ones from the MCMC
solutions in quadrature to obtain the final errors reported above in
Table~\ref{tab:lightcurves}. The derived quantities include, among
others, the individual Roche lobe radii as well as $r_{\rm vol}$, the
equivalent volume radius of each star (radius of a sphere with the
same volume as the Roche lobe).

%%%%%%%%%%%%%%%%%%%%%%%%%%%%%%%%%%%%%%%%%%%%%%%%%%%%%%%%%%%%%%%%%%%%%%
\section{Absolute dimensions}
\label{sec:dimensions}
%%%%%%%%%%%%%%%%%%%%%%%%%%%%%%%%%%%%%%%%%%%%%%%%%%%%%%%%%%%%%%%%%%%%%%

The combination of the spectroscopic elements in
Table~\ref{tab:specorbit} and the light elements in
Table~\ref{tab:lightcurves} yields the physical properties for the
system given in Table~\ref{tab:dimensions}. The absolute masses and
radii are determined with relative precisions of about 0.7\% each.
The averages of the measured projected rotational
velocities from the CfA and Fairborn spectra agree well with the
expected $v \sin i$ values for synchronous rotation (listed in the
table), within the errors.

\setlength{\tabcolsep}{3pt}  % tighten to make table fit in one column
\begin{table}
\centering
\caption{Physical Properties of \vstar. \label{tab:dimensions}}
\begin{tabular*}{0.475\textwidth}{lcc}
\hline \\ [-2.5ex]
\hline \\ [-2.5ex]
~~~~~~~~~~~Parameter~~~~~~~~~~~ & Primary & Secondary \\ [+0.5ex]
\hline \\ [-2ex]
 $M$ ($\mathcal{M}_{\sun}^{\rm N}$)\dotfill   &  $1.4153 \pm 0.0100$           &  $1.4023 \pm 0.0094$          \\ %\1ex]
 $R$ ($\mathcal{R}_{\sun}^{\rm N}$)\dotfill   &  $1.725 \pm 0.010$        &  $1.692 \pm 0.012$      \\ %\1ex]
 $q \equiv M_2/M_1$\dotfill               &          \multicolumn{2}{c}{$0.9909 \pm 0.0042$}          \\ %\1ex]
 $a$ ($\mathcal{R}_{\sun}^{\rm N}$)\dotfill                 &          \multicolumn{2}{c}{$6.182 \pm 0.013$}          \\ %\1ex]
 $\log g$ (cgs, dex)\dotfill              &  $4.1155 \pm 0.0061$           &  $4.1284 \pm 0.0067$          \\ %\1ex]
 $T_{\rm eff}$ (K)\dotfill                &  6840~$\pm$~150\phn            &  6780~$\pm$~110\phn \\ %\1ex]
 $L$ ($L_{\sun}$)\dotfill                 &  $5.84 \pm 0.52$               &  $5.42 \pm 0.36$           \\ %\1ex]
 $M_{\rm bol}$ (mag)\dotfill              &  $2.816 \pm 0.096$             &  $2.896 \pm 0.072$         \\ %\1ex]
 $BC_V$ (mag)\dotfill                     &  $+0.025 \pm 0.100$\phs        &  $+0.023 \pm 0.100$\phs \\ %\1ex]
 $M_V$ (mag)\dotfill                      &  $2.79 \pm 0.14$               &  $2.87 \pm 0.13$        \\ %\1ex]
 $v_{\rm sync} \sin i$ (\kms)\tablenotemark{a}\dotfill     &  $82.3 \pm 0.5$\phn            &  $80.7 \pm 0.6$\phn            \\ %\1ex]
 $v \sin i$ (\kms)\tablenotemark{b}\dotfill           &  $80.5 \pm 2.1$\phn       &  $80.5 \pm 2.1$\phn \\ %\1ex]
 $E(B-V)$ (mag)\dotfill                   &          \multicolumn{2}{c}{$0.088 \pm 0.020$} \\ %\1ex]
 $A_V$ (mag)\dotfill                      &          \multicolumn{2}{c}{$0.273 \pm 0.062$} \\ %\1ex]
 Dist.\ modulus (mag)\dotfill             &          \multicolumn{2}{c}{$8.76 \pm 0.12$} \\ %\1ex]
 Distance (pc)\dotfill                    &          \multicolumn{2}{c}{$564 \pm 30$\phn} \\ %\1ex]
% $\pi$ (mas)\dotfill                      &          \multicolumn{2}{c}{$1.77 \pm 0.10$} \\ %\1ex]
% $\pi_{Gaia/{\rm DR2}}$ (mas)\dotfill     &          \multicolumn{2}{c}{$1.788 \pm 0.036$} \\
 {\it Gaia}/DR2 distance (pc)\dotfill     &          \multicolumn{2}{c}{$559 \pm 11$\phn} \\ [+0.5ex]
\hline \\ [-2,5ex]
\multicolumn{3}{p{.47\textwidth}}{$^{\rm a}$ Synchronous projected rotational velocity assuming spin-orbit alignment.} \\
\multicolumn{3}{p{.47\textwidth}}{$^{\rm b}$ Average measured projected rotational velocity from CfA and Fairborn Observatory.} \\
\multicolumn{3}{p{.47\textwidth}}{{\bf Note}. --- The masses, radii, and semimajor axis $a$ are expressed
  in units of the nominal solar mass and radius
  ($\mathcal{M}_{\sun}^{\rm N}$, $\mathcal{R}_{\sun}^{\rm N}$) as
  recommended by 2015 IAU Resolution B3 \citep[see][]{Prsa:2016}, and
  the adopted solar temperature is 5772~K (2015 IAU Resolution
  B2). Bolometric corrections are from the work of \cite{Flower:1996},
  with conservative uncertainties of 0.1~mag, and the bolometric
  magnitude adopted for the Sun appropriate for this $BC_V$ scale is
  $M_{\rm bol}^{\sun} = 4.732$ \citep[see][]{Torres:2010b}. See text
  for the source of the reddening. For the apparent visual magnitude
  of \vstar\ out of eclipse we used $V = 11.11 \pm 0.02$
  \citep{Henden:2014, Henden:2015}. }
\end{tabular*}
\end{table}
\setlength{\tabcolsep}{6pt}  % back to default separation

Consistent estimates of the $E(B-V)$ reddening in the
direction of \vstar\ were obtained from five different sources:
0.083 \citep{Burstein:1982},
0.099 \citep{Drimmel:2003},
0.091 \citep{Amores:2005},
0.086 \citep{Schlafly:2011}, and
0.083 \citep{Green:2018}.
The straight average with a conservative uncertainty is $E(B-V) =
0.088 \pm 0.020$~mag, from which the extinction is $A_V = 3.1 E(B-V) =
0.273 \pm 0.062$ mag.

Using this value of $A_V$, the distance to the system was inferred
from radii and temperatures, the out-of-eclipse brightness of $V =
11.11 \pm 0.02$ \citep{Henden:2014, Henden:2015}, and bolometric
corrections from \cite{Flower:1996}, and is $564 \pm 30$~pc. This is
very nearly the same as the more precise distance of $559 \pm 11$~pc
inferred from the {\it Gaia}/DR2 parallax \citep{Gaia:2018}, and the
agreement speaks indirectly to the combined accuracy of our radii and
effective temperatures.

As an additional check on the spectroscopic temperatures, we collected
brightness measurements of the combined light of the binary from the
literature in the Johnson-Cousins and 2MASS systems
\citep{Droege:2006, Skrutskie:2006, Henden:2014, Henden:2015},
rejecting others that are known to have been taken in eclipse. We
constructed six non-independent color indices, corrected them for
reddening following \cite{Cardelli:1989}, and used color-temperature
calibrations by \cite{Casagrande:2010} to infer photometric
temperatures from each index. The weighted mean of the six values,
$6850 \pm 70$~K, is very close to the average of the spectroscopic
temperatures (6810~K), supporting the accuracy of those values.  The
temperature difference between the components is measured much more
precisely from the light curve analysis than from the CfA spectra, and
is $\Delta T_{\rm eff} = 59 \pm 24$~K.

%%%%%%%%%%%%%%%%%%%%%%%%%%%%%%%%%%%%%%%%%%%%%%%%%%%%%%%%%%%%%%%%%%%%%%
\section{Comparison with theory}
\label{sec:models}
%%%%%%%%%%%%%%%%%%%%%%%%%%%%%%%%%%%%%%%%%%%%%%%%%%%%%%%%%%%%%%%%%%%%%%

\begin{figure}
\epsscale{1.15}
\plotone{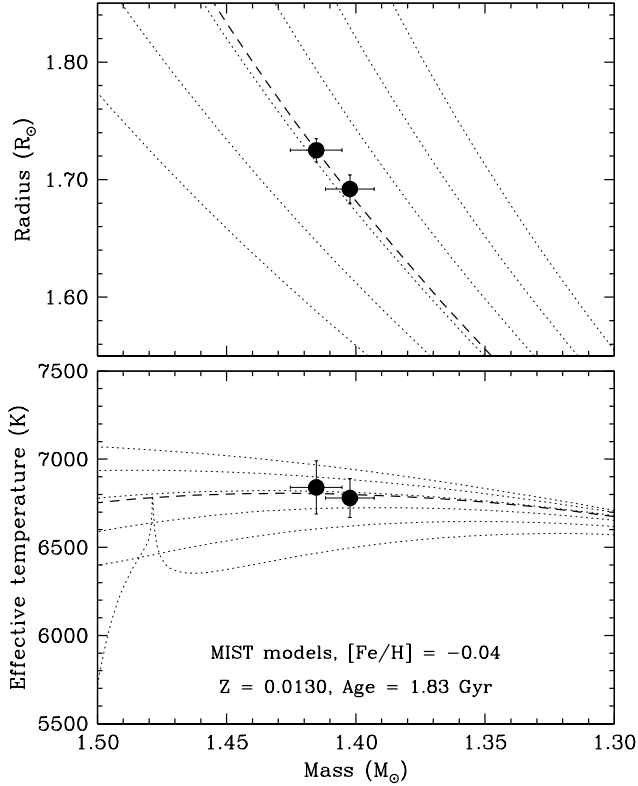}

\caption{Mass-radius and mass-temperature diagrams for \vstar\ showing
  isochrones from the MIST series \citep{Choi:2016} for the
  best-fitting metallicity of ${\rm [Fe/H]} = -0.04$. Dotted lines
  correspond to ages of 1.4--2.4~Gyr in steps of 0.2~Gyr, and the
  best-fit age of 1.83~Gyr is indicated with a heavier dashed
  line.\label{fig:massradteff}}

\end{figure}

The very precise absolute dimensions of \vstar\ offer an opportunity
to test current stellar evolution models. Mass-radius and
mass-temperature diagrams are shown in Figure~\ref{fig:massradteff},
in which the observations are compared against model isochrones from
the MESA Isochrones and Stellar Tracks series
\citep[MIST;][]{Choi:2016}, which is based on the Modules for
Experiments in Stellar Astrophysics package
\citep[MESA;][]{Paxton:2011, Paxton:2013, Paxton:2015}. To our
knowledge there is no spectroscopic determination available for the
metallicity of \vstar.  We find that a slight adjustment in the
metallicity of the models from solar to ${\rm [Fe/H]} = -0.04$
provides an excellent fit to both radii and both effective
temperatures at the measured masses. The age of the system according
to these models is 1.83~Gyr, which is shown by the thick dashed line
in Figure~\ref{fig:massradteff}.

\begin{figure}
\centering
\epsscale{1.15}
\plotone{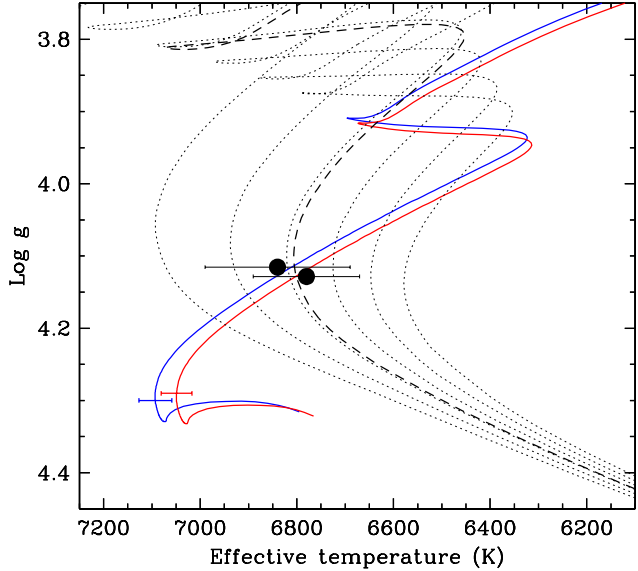}

\caption{Evolutionary trackes for the measured masses of the
  \vstar\ components and ${\rm [Fe/H]} = -0.04$. MIST isochrones
  \citep{Choi:2016} are shown with dotted lines for ages between 1.4
  and 2.4~Gyr, as in Figure~\ref{fig:massradteff}, with the best-fit
  age of 1.83~Gyr drawn as a thick dashed line. The uncertainty in the
  placement of the tracks that comes from the mass errors is indicated
  with the small error bars near the bottom of the
  tracks.\label{fig:tracks}}

\end{figure}

Evolutionary tracks for the measured masses are seen in
Figure~\ref{fig:tracks}, and indicate the components are halfway
through their main-sequence lifetimes. The uncertainty in the location
of the tracks due to the mass errors is shown at the bottom, and
corresponds to only about $\pm 30$~K in this diagram.

%%%%%%%%%%%%%%%%%%%%%%%%%%%%%%%%%%%%%%%%%%%%%%%%%%%%%%%%%%%%%%%%%%%%%%
\section{Discussion}
\label{sec:discussion}
%%%%%%%%%%%%%%%%%%%%%%%%%%%%%%%%%%%%%%%%%%%%%%%%%%%%%%%%%%%%%%%%%%%%%%

\vstar\ has been listed as a possible member of the sparse open
cluster Collinder~359 (Melotte~186) \citep{Sahade:1960, Sahade:1963},
although the location of the binary nearly 7\arcdeg\ from the cluster
center makes this rather unlikely a priori. Curiously, many of the
\vstar\ properties appear consistent with membership. For example, the
recent study by \cite{Cantat-Gaudin:2018} listed the parallax of
Collinder~359 as $\pi = 1.93 \pm 0.10$~mas, corresponding to a
distance of about $520 \pm 27$~pc, which is consistent with what we
obtain for the binary ($564 \pm 31$~pc; Table~\ref{tab:dimensions}).
\cite{Kharchenko:2005} reported the mean radial velocity of the
cluster to be $-4.45 \pm 0.25~\kms$, though based on measurements for
only two stars. This is also tantalizingly close to the center-of-mass
velocity we measured for \vstar, $-3.88 \pm 0.43~\kms$. The mean
proper motion components of Collinder~359 listed by
\cite{Cantat-Gaudin:2018} are $\mu_{\alpha} \cos\delta = +1.98 \pm
0.23$~mas~yr$^{-1}$ and $\mu_{\delta} = -8.19 \pm 0.25$~mas~yr$^{-1}$
based on the Fourth U.S.\ Naval Observatory CCD Astrograph Catalog
\citep[UCAC4;][]{Zacharias:2013}. Those of \vstar\ in the same catalog
are $\mu_{\alpha} \cos\delta = -1.8 \pm 1.4$~mas~yr$^{-1}$ and
$\mu_{\delta} = -4.9 \pm 1.5$~mas~yr$^{-1}$, which differ at about the
2.5$\sigma$ level from the cluster mean. However, if the $\sim$30~Myr
age of Collinder~359 reported by \cite{Kharchenko:2005} is accurate,
then \vstar\ cannot be a member, as we find it to be much older
(1.83~Gyr).

\vstar\ joins the ranks of the detached eclipsing binaries with the
very best determined properties \citep[see, e.g.,][]{Torres:2010a}.
Its value for testing models of stellar evolution would be
significantly enhanced by a spectroscopic determination of the
metallicity, although this may be challenging given the significant
line broadening of both stars.

\acknowledgments

We are grateful to the observers P.\ Berlind, M.\ Calkins, and
G.\ Esquerdo for their assistance in obtaining the CfA spectra.
J.\ Mink is also acknowledged for maintaining the CfA echelle data
base. The anonymous referee provided helpful comments that improved
the original manuscript. The authors wish to thank Bill Neely, who
operates and maintains the NFO WebScope for the Consortium, and who
handles preliminary processing of the images and their distribution.
GT acknowledges partial support from the NSF through grant
AST-1509375.  Astronomy at Tennessee State University is supported by
the state of Tennessee through its Centers of Excellence Program. The
computational resources used for this research include the Smithsonian
Institution's ``Hydra'' High Performance Cluster. This research has
made use of the SIMBAD database and the VizieR catalogue access tool,
both operated at the CDS, Strasbourg, France, and of NASA's
Astrophysics Data System Abstract Service.  This work has also made
use of data from the European Space Agency (ESA) mission {\it Gaia\/}
(\url{https://www.cosmos.esa.int/gaia}), processed by the {\it Gaia\/}
Data Processing and Analysis Consortium (DPAC,
\url{https://www.cosmos.esa.int/web/gaia/dpac/consortium}). Funding
for the DPAC has been provided by national institutions, in particular
the institutions participating in the Gaia Multilateral Agreement.

\end{document}